\documentstyle[epsfg]{aipproc}

\def\ltsima{$\; \buildrel < \over \sim \;$}
\def\lsim{\lower.5ex\hbox{\ltsima}}
\def\gtsima{$\; \buildrel > \over \sim \;$}
\def\gsim{\lower.5ex\hbox{\gtsima}}
\def\lcdm{$\Lambda$CDM}

\def\nwm2sr{nW m$^{-2}$ sr$^{-1}$}

\begin{document}

\title{Probing Galaxy Formation with High Energy Gamma-Rays}

\author{Joel R. Primack$^a$, Rachel S. Somerville$^b$, James S.
Bullock$^c$, and Julien E. G. Devriendt$^d$}
\address{$^a$Physics Department, University of California, Santa Cruz,
CA 95060 USA \\ 
$^b$ Institute of Astronomy, Madingley Rd., Cambridge CB3
OHA, UK \\
$^c$ Department of Astronomy, Ohio State University,
Columbus, OH 43210 USA \\
$^d$ Nuclear and Astrophysics Laboratory, Keble Road, OX1 3RH Oxford, UK}

\maketitle

\begin{abstract}
We discuss how measurements of the absorption of $\gamma$-rays from
GeV to TeV energies via pair production on the extragalactic
background light (EBL) can probe important issues in galaxy formation.
We use semi-analytic models (SAMs) of galaxy formation, set within the
hierarchical structure formation scenario, to obtain predictions of
the EBL from 0.1 to 1000$\mu$m. SAMs incorporate simplified physical
treatments of the key processes of galaxy formation --- including
gravitational collapse and merging of dark matter halos, gas cooling
and dissipation, star formation, supernova feedback and metal
production --- and have been shown to reproduce key observations at
low and high redshift.  Here we also introduce improved modelling of
the spectral energy distributions in the mid-to-far-IR arising from
emission by dust grains.  Assuming a flat \lcdm\ cosmology with
$\Omega_m=0.3$ and Hubble parameter $h=0.65$, we investigate the
consequences of variations in input assumptions such as the stellar
initial mass function (IMF) and the efficiency of converting cold gas
into stars. We conclude that observational studies of the absorption
of $\gamma$-rays with energies from $\sim$10 Gev to $\sim$10 TeV will
help to determine the EBL, and also help to explain its origin by
constraining some of the most uncertain features of galaxy formation
theory, including the IMF, the history of star formation, and the
reprocessing of light by dust.
\end{abstract}

\section{Introduction}

The extragalactic background light (EBL) represents all the light that
has been emitted by galaxies over the entire history of the
universe. The EBL that we observe today is an admixture of light from
different epochs, its spectral energy distribution (SED) distorted by
the redshifting of photons as they travel to us from sources at
different distances. It is therefore a constraint on both the
intrinsic SEDs of the sources and their distribution in redshift.  At
present, there is more than a factor of two uncertainty in the
amplitude of the EBL in the UV, optical, and near-infrared
\cite{puget}. The EBL in the mid-IR is even more uncertain. The far-IR
background measured at $\gsim 100 \mu$m
\cite{puget96,guider97,hauser,fixsen} represents at least half of the
total energy in the EBL, yet the sources that produced it remain
uncertain.

High energy $\gamma$-ray astronomy promises to help resolve these
uncertainties by providing independent constraints on the EBL, in the
mid-IR with $E_\gamma$ in the $\sim 10$ TeV energy range, and in the
0.1-3 $\mu$m range with $E_\gamma \sim100$ GeV via the new
low-threshold instruments that will soon be available.  High energy
$\gamma$-rays from sources at cosmological distances are absorbed via
electron-positron pair production on the diffuse background of photons
that comprises the EBL. Thus, $\gamma$-ray observations of objects
with known redshift and intrinsic spectral shape will constrain the
EBL in these crucial wavelength regimes by measuring the optical depth
of the Universe to photons of various energies.  This in turn will
help to constrain some of the most fundamental uncertainties in
physical models of galaxy formation.

In order to illustrate this, in this paper we use a ``forward
evolution'' approach, which attempts to model the essential features
of galaxy formation using simple recipes. These semi-analytic models
are set within the modern Cold Dark Matter (CDM) paradigm of
hierarchical structure formation, and trace the gravitational collapse
and merging of dark matter halos, the cooling and shock heating of
gas, star formation, supernovae feedback, metal production, the
evolution of stellar populations and the absorption and re-emission of
starlight by dust. This machinery has been used extensively to predict
optical properties of low-redshift galaxies, with good results (e.g.,
\cite{kwg,cole94}; reviewed and extended in \cite{sp,spf}, hereafter
SP and SPF).  A semi-analytic approach was also used by Devriendt and
Guiderdoni \cite{dev2} to make predictions of counts and backgrounds
in the mid-to-far-IR, with more detailed modelling of dust extinction
and emission, but less detailed modelling of merging and star
formation.  We have now combined the strengths of these two models, by
integrating the stellar SEDs and dust modelling of \cite{dev1,dev2}
into the galaxy formation code of the Santa Cruz group.

Some parts of the ``standard paradigm'' of galaxy formation
represented by our SAMs are relatively solid. For example, once a
cosmological model and power spectrum are specified, it is
straightforward to compute the gravitational collapse of dark matter
into bound halos using $N$-body techniques, and analytic formalisms
such as those used in our modelling \cite{sk} have been checked
against these results \cite{slkd}. Within the range of values for the
cosmological parameters allowed by existing observational constraints
(i.e., $\Omega_{\rm matter} \simeq 0.3-0.5$, $\Omega_{\rm
matter}+\Omega_\Lambda \simeq 1$, $H_0 \simeq 60-80$ km/s/Mpc; see
e.g. \cite{primack2000} for a summary), these results do not change
significantly.  Similarly, modelling of gas cooling appears to be
fairly robust and agrees well with hydrodynamic simulations
\cite{pearce}. However, other aspects, notably the efficiency of
conversion of cold gas into stars, the effect of subsequent feedback
due to supernovae winds or ionizing photons, the stellar initial mass
function (IMF), and the effects of dust, remain highly uncertain, and
some predictions are quite sensitive to their details.

For example, SPF showed that the star formation history of the
Universe and the number density of high redshift $z \gsim 2$
``Lyman-break'' galaxies (LBGs; e.g. \cite{steidel:99}) may be quite
different depending on whether star formation is primarily regulated
by internal properties, such as gas surface density in a quiescent
disk, or triggered by an external event such as an interaction.
Because the largest samples of LBGs are primarily identified in the
rest UV, model predictions are also quite sensitive to the
high-stellar-mass slope of the IMF, and to dust extinction. At the
other end of the spectrum is the sub-mm population detected by SCUBA,
believed to be predominantly high redshift ($z \gsim 2$) 
luminous and ultraluminous infrared galaxies (LIRGs and ULIRGs)
powered by star formation rates of hundreds to thousands of solar
masses per year (e.g., \cite{sanders}). Theoretical predictions of the
numbers and nature of these objects are highly sensitive to the same
issues (the dominant mode of star formation, dust, the IMF), but
provide a crucial counter-balance to the optical observations. 
However, the current mismatch between the sensitivity and spatial
resolution of optical and sub-mm instrumentation has made it difficult
to establish the connection between the two populations
observationally.

The Milky Way, like most nearby galaxies, emits the majority of its
light in optical and near-IR wavelengths; only about 30\% of the
bolometric luminosity locally is released in the far-infrared
\cite{sn:91}.  This was generally believed to be typical of most of
the starlight at all redshifts until the discovery of the far-IR part
of the EBL by the DIRBE and FIRAS instruments on the COBE satellite,
at a level ten times higher than the no-evolution predictions based on
the local luminosity function of IRAS galaxies, and representing twice
as much energy as the optical background obtained from counts of
resolved galaxies \cite{madaupoz}. This result suggests that either
the dust extinction properties of ``normal'' galaxies change
dramatically with redshift, or a population of heavily extinguished
galaxies (perhaps analogous to local LIRGs and ULIRGs) is much more
common at high redshift than locally, or both.  Some of these galaxies
may have already been observed, at 15 $\mu$m by ISO \cite{elbaz99},
and at 850 $\mu$m by SCUBA \cite{blain}.

Guiderdoni et al. \cite{ghbm,dev2} showed that their simplified
semi-analytic model could reproduce the multi-wavelength data only if
they introduced a population of heavily extinguished galaxies with
high star formation rates, and with strong evolution of number density
with redshift.  This population was introduced ad-hoc by
\cite{ghbm,dev2}, but as discussed by these authors, by \cite{silkdev}
(based on \cite{bss}), and also by SPF,
the increasing importance of starbursts at high redshift, due to the
increasing merger rate and higher gas fractions, is a natural
mechanism to produce this population. The models of SPF contain a
detailed treatment of mergers and the ensuing collisional starbursts,
which has been calibrated against the merger rate in cosmological
$N$-body simulations \cite{kolatt} and the starburst efficiency in
hydrodynamical simulations \cite{mh,somerville}.  Moreover, they
produced good agreement with observations of LBGs and Damped
Lyman-$\alpha$ systems (SPF) as well as low redshift galaxies
(SP). Therefore, it will be extremely interesting to see if these same
models, when combined with the more sophisticated treatment of dust
extinction and emission developed by Devriendt, Guiderdoni, and
collaborators, will be able to simultaneously reproduce observations
over the broad range of wavelengths and redshifts discussed above.

In the next section we briefly describe the ingredients of our models,
and then present the results of the predicted EBL.  The following
section presents the implications for $\gamma$-ray attenuation, and
the final one briefly discusses some alternative treatments and our
own conclusions.  The work summarized here is a brief, preliminary
sample of the results which will soon be presented in a series of
papers, now in preparation, on the EBL and its breakdown into various
kinds of sources and on the implications for $\gamma$-ray astronomy.

\section{Semi-analytic modelling}

\begin{figure} 
\noindent \begin{minipage}[t]{2.5in} \centering 
 \psfig{file=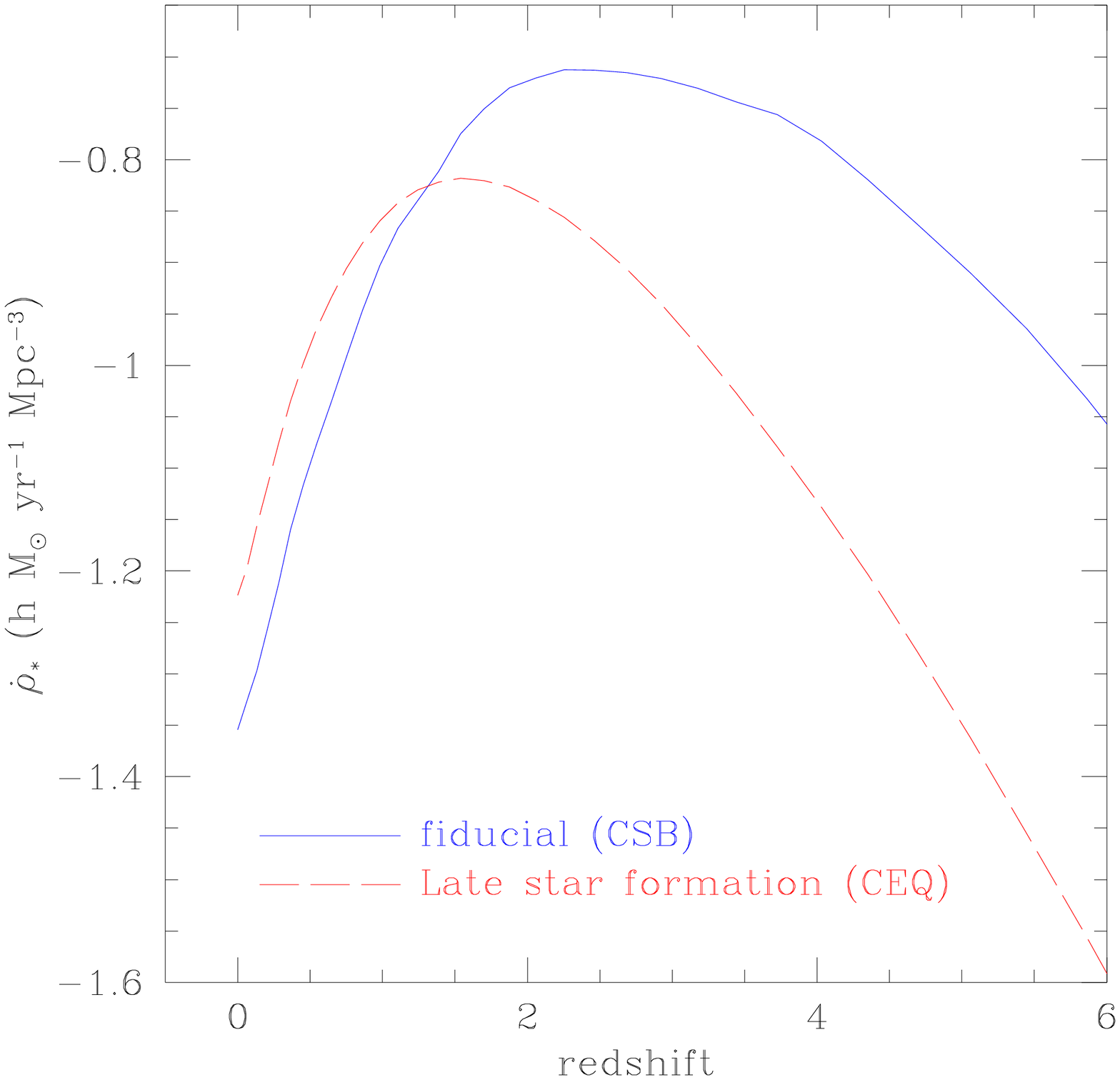,width=2.5in} \end{minipage} \hfill
\begin{minipage}[t]{2.5in} \centering 
 \psfig{file=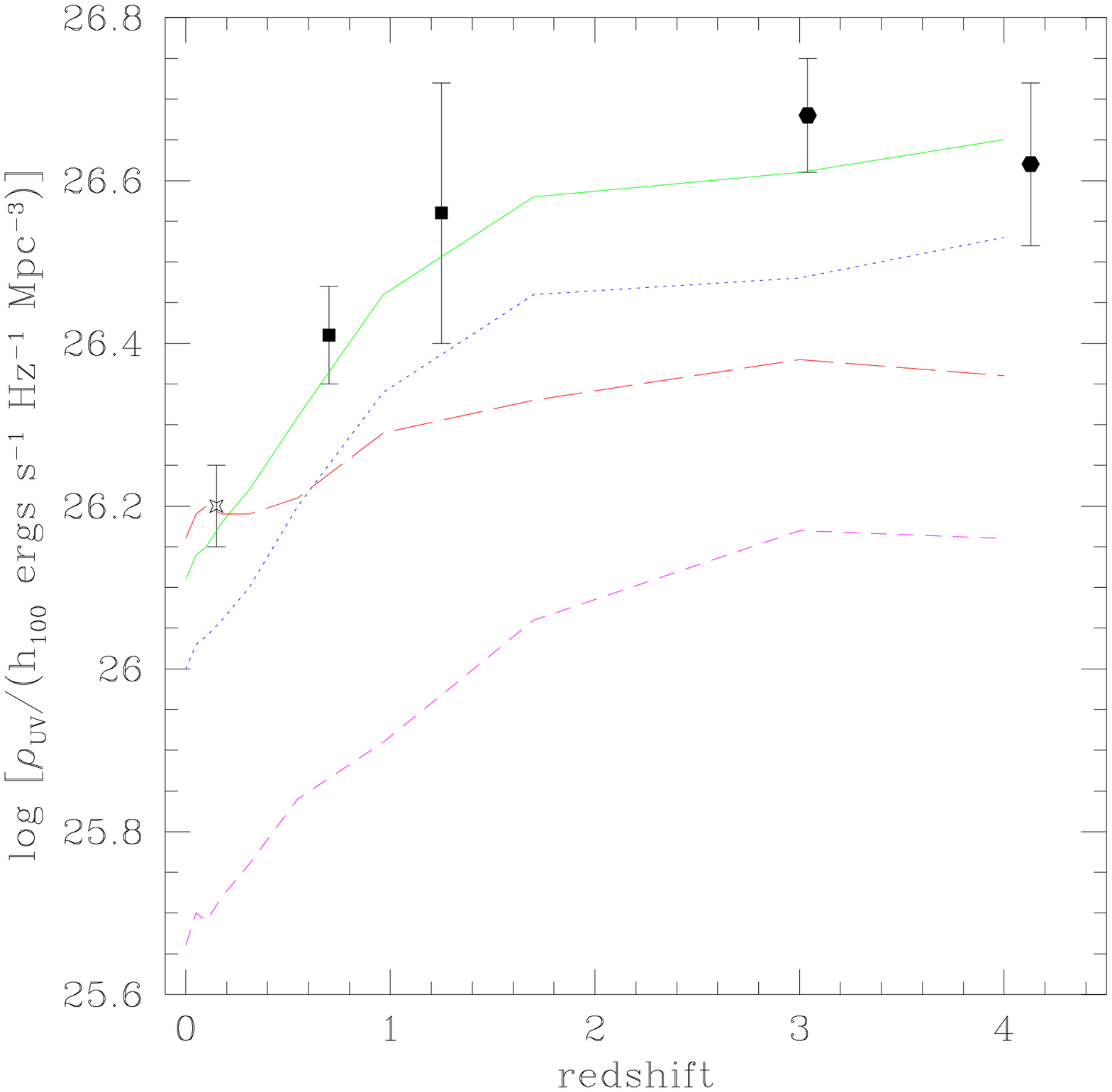,width=2.5in} \end{minipage} 
\caption{(a) The star formation rate density predicted by our models,
for two different recipes of star formation. Both models produce about
the same total mass density of stars by $z=0$ (i.e., the areas under
the curves are equal when they are plotted linearly vs. time), but the
collisional starburst model (CSB) peaks at higher redshift.  (b) Comoving
luminosity density at $2000 \AA$ as a function of redshift.  Data
points represent the observed global luminosity density at rest $\sim
2000 \AA$, obtained by integrating the observational best-fit
Schechter luminosity functions over all luminosities ($\rho_L = \phi_*
L_* \Gamma(2-\alpha)$). The $z=0.15$ point is from
\protect\cite{sullivan}, the $z\sim 0.4$ and 1.2 points are from
\protect\cite{cowie}, and the $z\sim3$ and $z\sim4$ points are from
\protect\cite{steidel:99}. The curves for our four models are labeled
as in Figure~\ref{fig:lumdens}.  The model curves have been corrected for
dust extinction using the approach described in the text.}
\label{UVlumdens}
\end{figure}

\begin{figure}[b!] 
\centerline{\epsfig{file=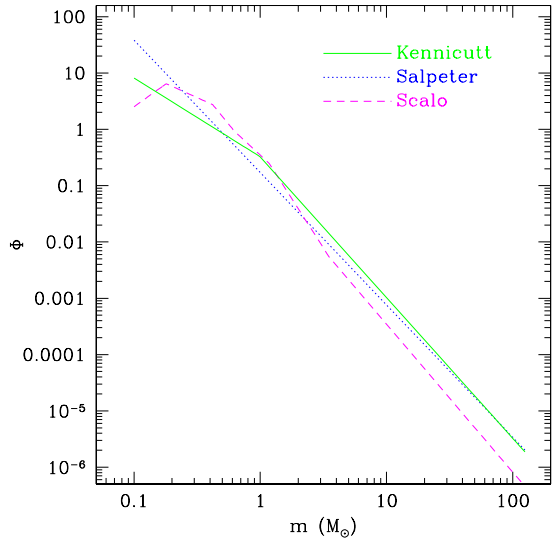,width=3in}}
\vspace{10pt}
\caption{The three stellar Initial Mass Functions (IMFs) used here:
Kennicutt \protect\cite{kenIMF}, Salpeter \protect\cite{salpeterIMF},
and Scalo \protect\cite{scaloIMF}.
}
\label{fig:imf}
\end{figure}

In this section we briefly describe the ingredients of our
models. Readers can refer to SP and SPF for more details, and to
\cite{veritas} for a brief introduction.

Using the method described in \cite{sk}, we create Monte-Carlo
realizations of the masses of progenitor halos and the redshifts at
which they merge to form a larger halo. These ``merger trees'' (each
branch in the tree represents a halo merging event) reflect the
collapse and merging of dark matter halos within a specific cosmology.
We truncate the trees at halos with a minimum circular velocity of 40
km/s, below which we assume that the gas is prevented from collapsing
and cooling by photoionization.  Each halo at the top level of the
hierarchy is assumed to be filled with hot gas, which cools
radiatively and collapses to form a gaseous disk. The cooling rate is
calculated from the density, metallicity, and temperature of the
gas. Cold gas is turned into stars using a simple recipe, depending on
the mass of cold gas present and the dynamical time of the
disk. Supernovae inject energy into the cold gas and may expell it
from the disk and/or halo if this energy is larger than the escape
velocity of the system. Chemical evolution is traced assuming a
constant yield of metals per unit mass of new stars formed. Metals are
initially deposited into the cold gas, and may later be redistributed
by supernovae feedback, and mixed with the hot gas or the diffuse
(extra-halo) IGM.

\begin{figure}[t!] 
\centerline{\epsfig{file=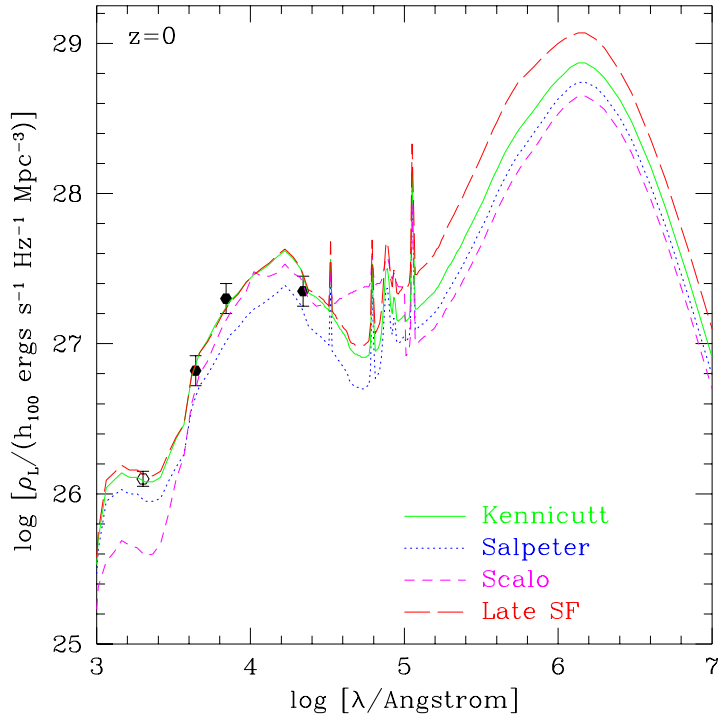,width=4in}}
\vspace{10pt}
\caption{Comoving luminosity density at redshift $z=0$ as a function
of wavelength. Data points represent the observed global luminosity
density of the Universe, obtained by integrating the observational
best-fit Schechter luminosity functions over all luminosities ($\rho_L
= \phi_* L_* \Gamma(2-\alpha)$). The far-UV point (at $\lambda=2000
\AA$) is from the luminosity function from FOCA observations
\protect\cite{sullivan}, extrapolated from the mean redshift of the
sample ($z=0.15$) to $z=0$ assuming that the luminosity density scales
with redshift as $\rho_L \propto (1+z)^{1.7}$, as indicated by the
observations of Cowie et al. \protect\cite{cowie}. The B-band point
($\lambda=4400\AA$) is from the luminosity function derived from the
2dF redshift survey \protect\cite{folkes}, the R-band point
($\lambda=6940\AA$) is from the Century Redshift Survey
\protect\cite{geller}, and the K-band point ($\lambda=2.2 \mu m$) is
from \protect\cite{gardnerK}. The model curves are obtained by simply
summing the spectra of all $z=0$ galaxies in our models with the
appropriate Press-Schechter weighting.}
\label{fig:lumdens}
\end{figure}

When halos merge, the galaxies contained in each progenitor halo
retain their seperate identities until they either spiral to the
center of the halo due to dynamical friction and merge with the
central galaxy, or until they experience a binding merger with another
satellite galaxy orbiting within the same halo.  All newly cooled gas is
assumed to initally collapse to form a disk, and major (nearly equal
mass) mergers result in the formation of a spheroid. New gas accretion
and star formation may later form a new disk, resulting in a variety
of bulge-to-disk ratios at late times.

For an assumed IMF, the stellar SED of each galaxy is then obtained
using stellar population models. Here we use the multi-metallicity
stellar SEDs of \cite{dev1} for the Salpeter and Kennicutt IMF cases,
and the solar metallicity GISSEL models \cite{bc} for the Scalo IMF.
(We have found that using evolving metallicity rather than solar
metallicity SEDs has a relatively small impact on the resulting EBL.)
Dust extinction is modelled using an approach similar to that of
\cite{dev2}. The optical depth of the disk is assumed to be
proportional to the column density of metals.
We then use a simple slab geometry where stars and gas are homogenously
mixed, and assign a random inclination to each galaxy to compute the
absorption. We use a metallicity dependent extinction curve, following
\cite{ghbm,dev2}.

All absorbed light is re-radiated at longer wavelength. In general,
any galactic dust emission spectrum can by represented by a
combination of three components: 1) hot dust (as in $H_{\rm II}$
regions), 2) warm dust (as in the diffuse $H_{\rm I}$), and 3) cold
dust (as in molecular clouds). In the models of Devriendt et
al. \cite{dev1}, these components are modelled as a mixture of
polycyclic aromatic hydrocarbon molecules (PAH), very small grains,
and big grains. Big grains may be either cold ($\sim$ 17 K), or heated
by radiation from star-forming regions (as suggested by observations
of typical local starburst galaxies like M82). A set of template
spectra is then constructed for galaxies of varying IR luminosity,
with admixtures of the various components selected in order to
reproduce the observed relations between IR/sub-mm color and IR
luminosity. A similar approach was used by \cite{dwek:98}, using a
mixture of a typical Orion-like $H_{\rm II}$ spectrum and an $H_{\rm
I}$ spectrum constructed to fit DIRBE observations of the diffuse ISM
\cite{dwek:97}. Here, we use the more empirical emission templates of
\cite{dwek:98} (kindly provided in electronic form by E. Dwek), but we
obtain very similar results with the models of \cite{dev1}.

The recipes for star formation, feedback, chemical evolution, and dust
optical depth contain free parameters, which we set for each model
(see SP) by requiring an average fiducial ``Milky Way'' galaxy to have
a K-band magnitude, cold gas mass, metallicity, and average B-band
extinction as dictated by observations of nearby galaxies.

Figure 1a shows the global star formation rate density for the two
star formation recipes that we consider here. The ``fiducial'' model
is the collisional starburst (CSB) model favored by SPF, in which
bursts of star formation may be triggered by galaxy collisions. The
``Late star formation'' model is the Constant Efficiency Quiescent
(CEQ) model of SPF, in which cold gas is converted to stars only in a
quiescent mode with constant efficiency. This produces a star
formation history similar to the models of \cite{baugh}, in which the
peak in the star formation history occurs at a considerably later
epoch ($z \sim 1.5$) than in the CSB model.  Figure~\ref{fig:lumdens}
shows the resulting luminosity density as a function of wavelength at
$z=0$. For the CSB model, we consider three different choices of IMF:
Scalo \cite{scaloIMF}, Salpeter \cite{salpeterIMF}, and Kennicutt
\cite{kenIMF}. These IMFs are graphed in Figure~\ref{fig:imf}.  For
the CEQ model we show only the Kennicutt case.  This is compared with
the observed luminosity density from nearby galaxies, obtained by
integrating the luminosity functions of galaxies resolved in recent
redshift surveys at wavelengths ranging from 0.2 to 2.2 $\mu$m.  The
spikes in the model predictions at $\sim 5-12 \mu$m are caused by the
PAH features mentioned above. All four models, when normalized to the
observed K-band Tully-Fisher relation, produce reasonable agreement
with the observed luminosity density in the B and K bands.\footnote{In
\cite{veritas}, we renormalized all the models by requiring that they
all agreed with the K-band point at 2.2 $\mu$m.  Here we do not do
this since, as Fig.~\ref{fig:lumdens} shows, the SAM parameters chosen
for each case to produce an average fiducial ``Milky Way'' as
described above are already in agreement with this data within the
errors.  Also, our current SAMs \protect\cite{sp,spf} use a corrected
version \protect\cite{shethtormen} of the Press-Schechter formalism,
which obviates our previous motivation for the K-band normalization.}
This is perhaps not surprising, yet it was not guaranteed. However,
there is a noticable difference in the far-UV and the mid- to far-IR.
The Scalo IMF produces too little UV light relative to optical and
near-IR light, whereas the Kennicutt and Salpeter IMFs are in much
better agreement with the data. These IMFs produce more high mass
stars than the Scalo IMF, and thus more ultraviolet light to be
absorbed and re-radiated by dust in the far IR. In Fig.~1b we show the
redshift evolution of the far-UV ($2000 \AA$) luminosity density,
compared with observations. The Scalo model falls short at all
redshifts, and the CEQ model, which agrees at $z=0$, falls short at
higher redshifts. It is encouraging that our very simple model for
dust extinction, which we normalized in the B-band at $z=0$, appears
to yield the appropriate level of dust extinction in the UV at higher
redshifts (SPF).

\begin{figure}[t!] 
\centerline{\epsfig{file=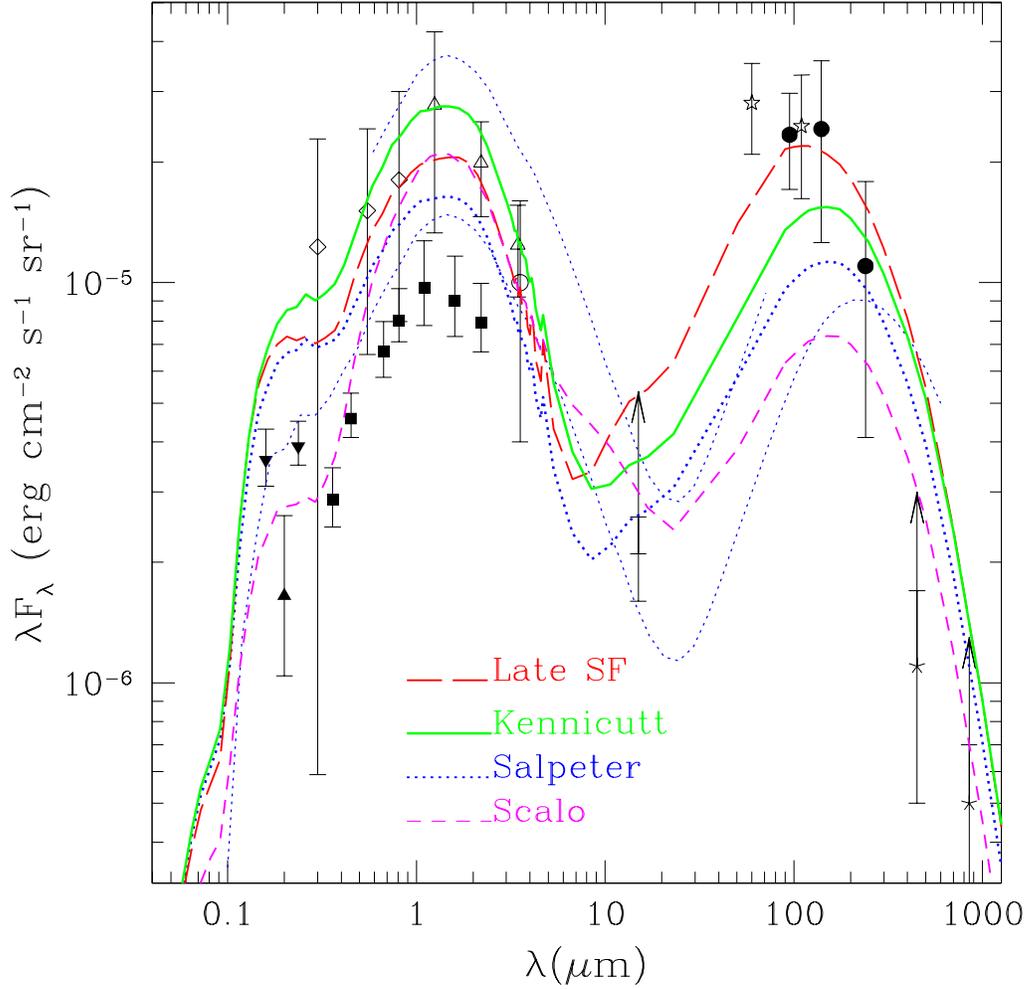,width=5.5in}}
\vspace{10pt}
\caption{Extragalactic background light: models and data. The far-UV
points are from STIS (inverted filled triangles)
\protect\cite{gardner} and FOCA observations (filled triangle)
\protect\cite{foca}.  The lower optical points (filled squares) are
lower limits from resolved sources \protect\cite{madaupoz}; the upper
ones (open diamonds) are from absolute photometry
\protect\cite{bernstein}.  The near-IR points are from DIRBE: (open
circle) \protect\cite{dwekarendt}, (open triangles)
\protect\cite{wright}.  The point at 15 $\mu$m is from ISOCAM resolved
sources \protect\cite{elbaz99}, and is thus a lower limit.  The far-IR
points are from DIRBE (filled circles) \protect\cite{hauser,lagache},
(stars) \protect\cite{fds}.  The curves are our results from modelling
the history of star formation in the \lcdm\ cosmology using
semi-analytic methods: a model with both quiescent star formation with
constant efficienty and starbursts, with Kennicutt, Salpeter, and
Scalo IMFs, and a model with only quiescent star formation with
constant efficienty (Late SF).  The lower light dotted curve is the
\lcdm\ EBL calculated using our previous methods
\protect\cite{veritas} for the Salpeter IMF, and the upper one is the
same curve to 80 $\mu$m multiplied by 2.5 for comparison with Mrk 501
data as analyzed by \protect\cite{guy} (see text).  Note that
$10^{-6}$ erg s$^{-1}$ cm$^{-2}$ sr$^{-1}$ = 1 \nwm2sr. }
\label{fig:ebl}
\end{figure}

\section{The Integrated Extragalactic Background Light}

Figure~\ref{fig:ebl} shows the EBL produced by our four models,
obtained by integrating the light over redshift (out to $z=4$) with
the appropriate K-corrections due to cosmological redshifting. We
compare this with a compilation of observational limits and
measurements of the EBL.  It is apparent that there is at least as
much energy in the far-IR part of the EBL as in the entire optical and
near-IR bands.  For example, Puget and collaborators \cite{puget}
estimated that the total energy in the EBL is between 60 and 93
\nwm2sr, with between 20 and 41 \nwm2sr\ contributed by the optical
and near-IR, and between 40 and 52 \nwm2sr coming from the far-IR.  If
the possible detection of the EBL at 60 $\mu$m by Finkbeiner et
al. \cite{fds} is correct, that would further increase the far-IR EBL;
however, as Puget discussed in his talk at this conference, it is very
difficult to determine the EBL at 60 $\mu$m since the zodiacal light
is so much brighter at that wavelength.

In units of critical density $\rho_c$, $\Omega_{\rm EBL} = (4\pi/c)
(I_{\rm EBL}/\rho_c c^2) = 2.5\times10^{-8} I_{\rm EBL} h^{-2}$, where
$I_{\rm EBL}$ is in units of \nwm2sr.  The total energy density in the
EBL corresponding to the lower and upper estimates of \cite{puget} is
$\Omega_{\rm EBL} = (3.6-5.5) \times 10^{-6} (h/0.65)^{-2}$.  Although
the EBL includes energy radiated by active galactic nuclei (AGNs) as
well as stars, it is unlikely that AGNs contributed more than a few
percent of the total.  The total energy radiated by AGNs is $E_{\rm
EBL}^{\rm AGN} = \eta \rho_{\rm BH} c^2$, where the efficiency of
conversion of mass to radiated energy in AGNs is $\eta \sim 0.05$.
Correspondingly, $\Omega_{\rm EBL}^{\rm AGN} = \eta \Omega_{\rm BH}
(1+z_{\rm BH})^{-1} \approx 4.5 \times 10^{-8} h^{-1} (\eta/0.05)
[3/(1+z_{\rm BH})] \lsim 0.02 \Omega_{\rm EBL}$.~\footnote{Updating
\cite{madau}, we have estimated $\Omega_{\rm BH}= (M_{\rm BH}/M_{\rm
spheroid}) \Omega_{\rm spheroid} \approx (1.5\times10^{-3})(1.8
\times10^{-3} h^{-1})$, using the observed (loose) correlation
\cite{kor} between a black hole mass and that of the galactic spheroid
in which it is found, and the estimated cosmological density of
spheroids \cite{fuku}.}  So for simplicity, in this paper we will
neglect the contribution of AGNs to the EBL.

Several interesting features emerge from the comparison of our SAM
models with the EBL data. In the UV to near-IR, the models are much
closer to the direct measures of the EBL obtained by
\cite{bernstein,dwekarendt,wright}, although the Scalo IMF produces
less light in the UV because it has fewer high-mass stars.  Recall
that the Kennicutt model agreed well with the observed luminosity
density at $z=0$, and the observed redshift evolution of the
luminosity density in the rest UV. This suggests that the extra factor
of 2-3 in the direct measurements of the EBL must arise from a rapidly
evolving population of objects which are too faint or too low in
surface brightness to be detected in the samples used to obtain the
counts (e.g., \cite{madaupoz}).  We are in the process of attempting
to determine whether the observational selection effects inherent in
the measured counts are sufficient to explain this discrepancy for our
modelled population. A second interesting point is that all the models
satisfy the lower limits from counts in the mid-IR (15 $\mu$m;
\cite{elbaz99}) and the sub-mm (850 $\mu$m; \cite{blain}).  Of our
four new EBL curves, the Late SF model and the fiducial Kennicutt
model are also consistent with the DIRBE/FIRAS measurements at 140 and
240 $\mu$m.  The models differ significantly in the mid-IR,
$\sim10-60\mu$m, where the EBL can be probed by TeV $\gamma$-rays.
The lower dotted curve, representing our previous attempt
\cite{veritas} to model the EBL, is well below the 15 $\mu$m lower
limit as well as the DIRBE measurements at longer wavelengths.  As we
stated in \cite{veritas}, we expected our EBL results to change as we
improved our dust emission modelling.  In addition to inclusion of the
PAH features, the new dust emission model has more warm dust than the
one used in \cite{veritas}.

\begin{figure}[b!] 
\centerline{\epsfig{file=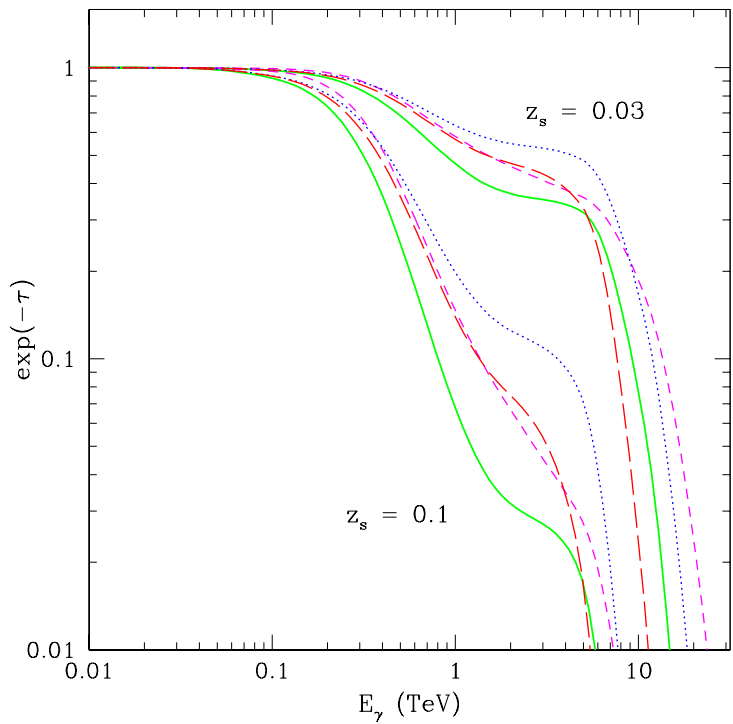,width=4in}}
\vspace{10pt}
\caption{The attenuation factor, $\exp(-\tau)$ for $\gamma$-rays as a
function of $\gamma$-ray energy for the four \lcdm\ models
considered in Fig. 4.  The assumed redshift of the source, z$_s$, is
indicated for each set of curves. 
}
\label{fig:atten}
\end{figure}

We now discuss constraints from the TeV $\gamma$-ray observations.

\section{Attenuation of high-energy $\gamma$-rays}

Figure~\ref{fig:atten} shows the $\gamma$-ray attenuation predicted by
the four \lcdm\ models considered here, for sources at redshifts
$z_s=0.03$ and 0.10.  All of the models predict rather little
absorption at $E_\gamma \lsim 5$ TeV for sources at $z_s=0.03$, but
fairly sharp cutoffs above 10 TeV, especially for the Late SF model.
That model may be in conflict with the data from Mrk 501.  The
synchrotron self-Compton (SSC) model, in which $\sim$keV synchrotron
X-radiation from a very energetic electron beam is Compton
up-scattered by the same electrons to produce the observed $\sim$TeV
$\gamma$-rays, appears to explain both the keV-TeV spectra and their
time variation for the blazars Mrk 421 and 501, both at $z\approx
0.03$ (see, e.g., \cite{guy,kraw99} and references therein).  Using a
simplified SSC model and keV X-ray data to predict the unattenuated
TeV spectrum of Mrk 501, Guy et al. \cite{guy} used CAT and HEGRA data
to estimate the amount of $\gamma$-ray attenuation.  They find that
there is a rather good fit to the observed attenuation for the
\lcdm-Salpeter EBL from \cite{veritas} when it is scaled upward by a
factor of up to about 2.5 across the wavelength range 1-80 $\mu$m;
this is the upper Salpeter curve on Fig.~\ref{fig:ebl}.  The
\cite{guy} 1-$\sigma$ upper limit for 20-80 $\mu$m is a scaling factor
of 3.4.  Our new Salpeter curve appears to be rather consistent with
this rescaling of our old Salpeter one, the Kennicutt curve may be a
little high, but the Late SF curve appears to be definitely too high.
\begin{figure}[t!] 
\centerline{\epsfig{file=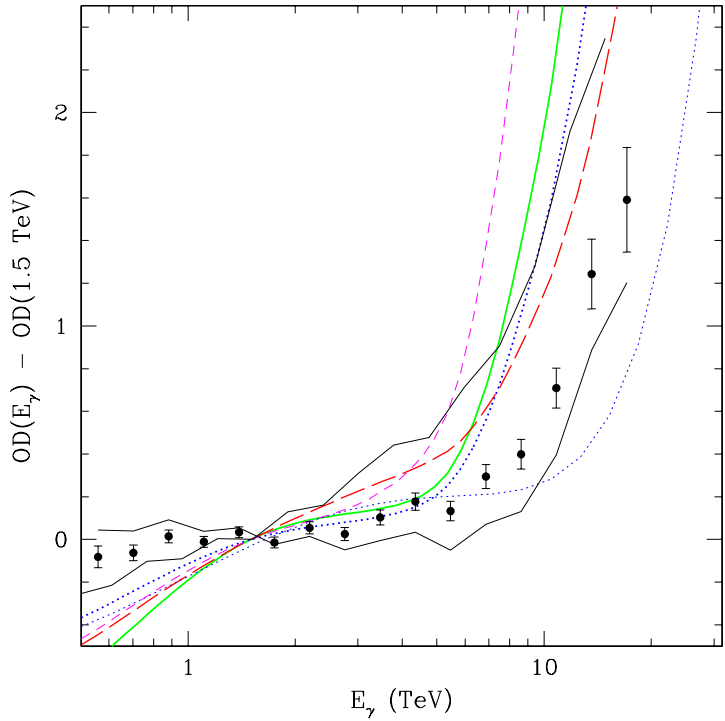,width=4in}}
\vspace{10pt}
\caption{Increase of the optical depth (OD) for $\gamma$-rays due to
intergalactic extinction, inferred from comparison of the observed Mrk
501 spectrum from HEGRA with that estimated using SSC models.  The
points correspond to Doppler factor $\delta_j=25$, magnetic field
$B=0.037$ G; the upper and lower light solid curves correspond
respectively to $(\delta_j,B)=(100,0.012$ G) and (25,0.12 G), and
the statistical error bars on the points also apply to these curves.
The redshift of Mrk 501 z$_s=0.034$ was used in calculating the
optical depth for each model.  This figure is based on Figure 10 of
\protect\cite{kraw99}. 
}
\label{fig:kraw}
\end{figure}

\begin{figure}[t] 
\centerline{\epsfig{file=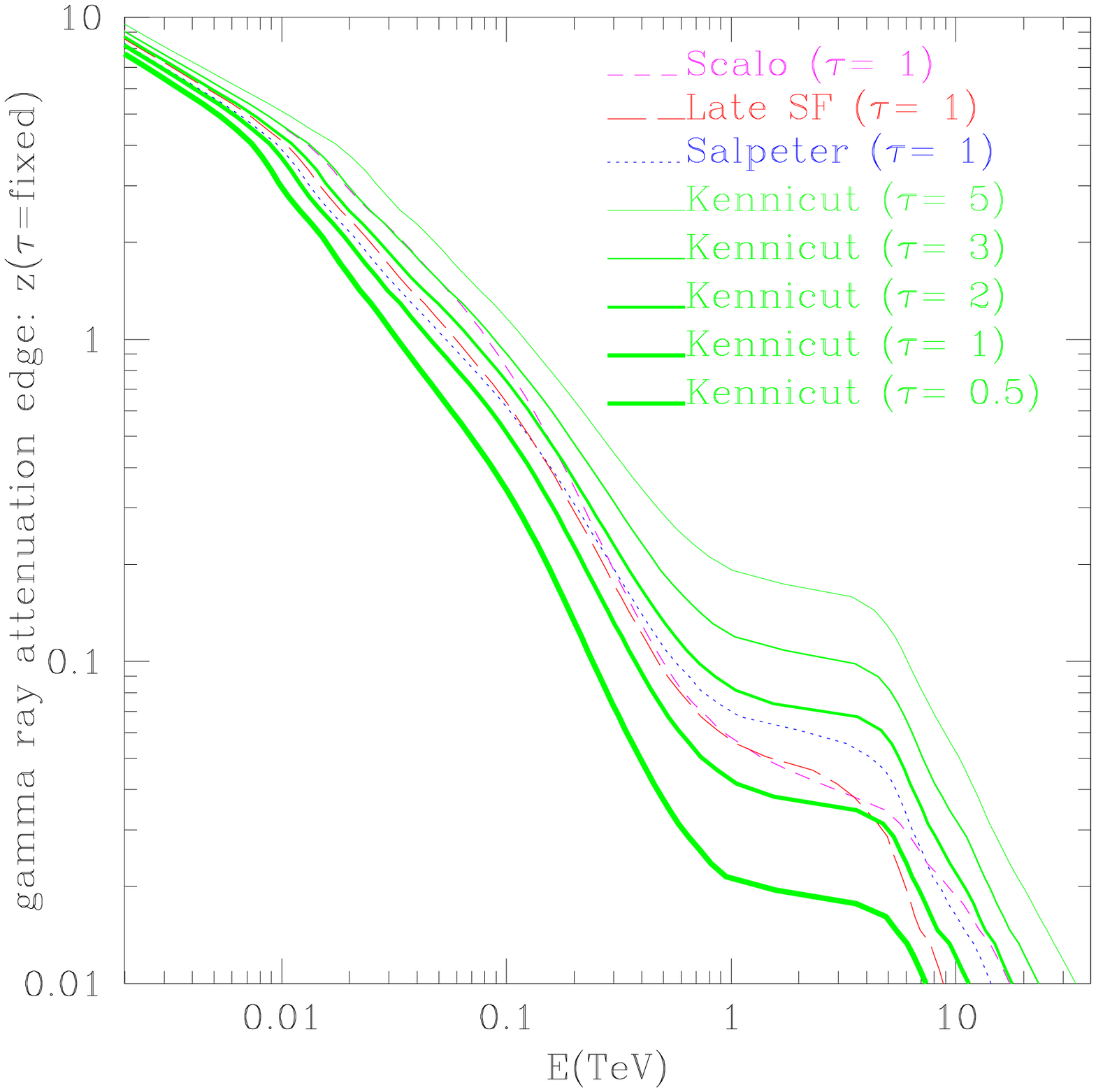,width=4in}}
\vspace{10pt}
\caption{The $\gamma$-ray attenuation edge.  The redshift where the
optical depth reaches unity is shown as a function of $\gamma$-ray
energy for each of the four \lcdm\ models considered in Fig. 4.  Also
shown for the Kennicutt IMF is the redshift where the optical depth
equals 0.5, 2, 3, and 5.}
\label{fig:edge}
\end{figure}
The compatibility of our new EBL calculations with the available data
on TeV $\gamma$-ray attenuation is definitely worth further
investigation.  The results appear to be sensitive to the details of
the models, raising the hope that they may be able to help answer
important questions about star formation and dust reradiation, and
also help to test the SSC modelling.  For example,
Figure~\ref{fig:kraw} shows the optical depth as a function of
$\gamma$-ray energy $E_\gamma$, compared with $\gamma$-ray attenuation
results from Mrk 501 with detailed SSC models.  This figure is like
Fig. 10 of Krawczynski et al. \cite{kraw99}.  While that figure showed
OD($E_\gamma$) - OD(0.5 TeV), following Krawczynski's advice we here
plot OD($E_\gamma$) - OD(1.5 TeV), normalizing to the data at 1.5 TeV
since the systematic error in the curvature of the spectrum strongly
increases below 1.5 TeV, corresponding at 0.5 TeV to a flux
uncertainty of 50\%.  (Krawczynski also kindly updated his model
curves for this figure to take into account the 15\% HEGRA energy
uncertainty.  We followed his advice to omit the highest-energy point,
which has a statistical significance well below 2$\sigma$.)  The
conclusions from Fig.~\ref{fig:kraw} appear to be consistent with
those from our discussion of \cite{guy}: the higher model curve seems
compatible with our new Salpeter results, taking into account that the
error bars on the data points also apply to the model curves; the
Kennicutt model also appears to be consistent, except perhaps at the
highest $E_\gamma$; and the Late SF model definitely predicts too much
attenuation.

Figure~\ref{fig:edge} depicts the $\gamma$-ray ``absorption edge,''
the redshift of a source corresponding to an optical depth of unity,
as a function of $\gamma$-ray energy.  Travelling through the evolving
extragalactic radiation field, $\gamma$-rays from sources at lower
redshift suffer little attenution.  The universe becomes increasingly
transparent as $E_\gamma$ decreases, probing the background light at
increasingly short wavelengths.  (We are using the treatment of
\cite{madau95} to account for absorption of ionizing radiation by the
Lyman alpha forest.)  The models all have the same qualitative
features, but differ significantly quantitatively.  The location of
the absorption edge is affected both by the assumed IMF and by the
history of star formation.  There is more absorption at most redshifts
with the Kennicutt IMF because with a higher fraction of high mass
stars, it is more efficient at producing radiation for a given stellar
mass; there is more absorption nearby in the Late SF model because the
starlight in this model is less diluted by the expansion of the
universe.  It is possible that measuring the transparency of the
universe to $\gamma$-rays at $\sim 0.1$ TeV with a number of sources
at various redshifts can provide a strong probe of star formation,
although there are uncertainties due to extinction by dust.

\section{Outlook}

The semi-analytic modelling of the EBL described here follows the
evolution of galaxy formation in time.  Forward modelling is a more
physical approach than backward modelling (luminosity evolution).
Pure luminosity evolution (e.g., \cite{ms98,ms00,steckeriau}) assumes
that the entire evolution of the luminosity of the universe arises
from galaxies in the local universe just becoming brighter at higher
redshift by some power of $(1+z)$ out to some maximum redshift.  It
effectively assumes that galaxies form at some high redshift and
subsequently just evolve in luminosity in a simple way.  This is at
variance with hierarchical structure formation of the sort predicted
by CDM-type models, which appears to be in better agreement with many
sorts of observations.

An alternative approach to modelling the EBL has been followed by Pei
and collaborators \cite{peifall95,fallcp,pei99}, in which they find an
overall fit to the global history of star formation subject to
constraints from input data including the evolution of the amount of
neutral hydrogen in damped Ly$\alpha$ systems (DLAS).  Their first
attempt \cite{peifall95,fallcp}, which was used as the basis for EBL
estimates by \cite{dwek:98,salamonstecker}, was somewhat misled by the
sharp drop in the DLAS hydrogen abundance from redshift $z\sim3$ to
$z\sim2$ reported in \cite{lwt95}.  With more complete data on DLAS
(see, e.g., Fig. 14 of \cite{slw00}) the $z=3$ neutral hydrogen
abundance is lower and almost constant from $z=2$ to 4.  The latest
paper by Pei et al. \cite{pei99} takes a variety of recent data into
account.  Their approach is to follow the evolution of the total mass
in stars, interstellar gas, and metals in a representative volume of
the universe; they assume a Salpeter IMF.  By contrast, the
semi-analytic methods we use follow the evolution of many individual
galaxies in the hierarchically merging halos of specific CDM models,
here \lcdm.  Despite the differences in approach, and the fact that
\cite{pei99} assumed $\Omega_m=1$ and Hubble parameter $h=0.5$, their
results are broadly similar to those from the semi-analytic approach
(see their \S4.4; for our semi-analytic approach to modelling DLAS,
see \cite{maller}).  In particular, their EBL is similar to our old
results \cite{veritas} for the Salpeter IMF.  Our EBL results
presented here are higher in the near-IR and more consistent with the
direct determinations \cite{dwekarendt,wright}; they are also higher in
the mid-IR, probably mainly because of the warm dust and PAH features
in our dust emission model.  It will be interesting to see whether
further development of the global approach of Pei et al. and of the
semi-analytic approach lead to convergent results.

As our calculations show, the EBL, especially at $\lsim 1$ $\mu$m and
$\gsim 10$ $\mu$m, is significantly affected by the IMF and the
absorption of starlight and its reradiation by dust, as well as by the
underlying cosmology.  The cosmological parameters are becoming
increasingly well determined by other observations.  As data become
available on $\gamma$-ray emission and absorption from sources at
various redshifts, especially from the new generation of Atmospheric
Cherenkov Telescopes and the new $\gamma$-ray satellites AGILE and
GLAST, these data and their theoretical interpretation will help to
answer fundamental questions concerning how and in what environments
all the stars in the universe formed.

\section*{Acknowledgments}

JRP was supported by NASA and NSF grants at UCSC, RSS by a rolling
grant from PPARC, and JSB by NASA LTSA grant NAG-3525 and NSF
AST-9802568.  JRP is grateful for a Humboldt Award, and he thanks Leo
Stodolsky for hospitality and Eckart Lorenz for enlightening
discussions about $\gamma$-ray astronomy at the Max-Planck-Institut
f\"ur Physik, M\"unchen.  We thank Eli Dwek for sending us his model
dust emission templates electronically, Henric Krawczynski for
updating his model points and curves for Fig.~\ref{fig:kraw} and for
very helpful discussions of his SSC modelling, and Felix Aharonian for
further discussion of SSC modelling and sage editorial suggestions.

\end{document}